\documentclass[twocolumn,pre,superscriptaddress]{revtex4}

\usepackage{graphics}
\usepackage{graphicx}
\usepackage{amsfonts}
\usepackage{textcomp}
\usepackage{amssymb}
\usepackage{mathrsfs}
\usepackage{amsmath}
\usepackage{color}
\usepackage{ulem}

\begin{document}

\title{Modeling Multi-Cellular Dynamics Regulated by ECM-Mediated Mechanical Communication via Active Particles with Polarized Effective Attraction}

\author{Yu Zheng\footnote{These authors contributed equally to this work.}}
\affiliation{Department of Physics, Arizona State University,
Tempe, AZ 85287}
\author{Qihui Fan\footnotemark[1]}
\affiliation{Beijing National Laboratory for Condensed Matte
Physics and CAS Key Laboratory of Soft Matter Physics, Institute
of Physics, Chinese Academy of Sciences, Beijing 100190, China}
\affiliation{School of Physical Sciences, University of Chinese
Academy of Sciences, Beijing 100049, China}
\author{Christopher Z. Eddy} \affiliation{Department of Physics, Oregon
State University, Corvallis, OR 97331}
\author{Xiaochen Wang}
\affiliation{Beijing National Laboratory for Condensed Matte
Physics and CAS Key Laboratory of Soft Matter Physics, Institute
of Physics, Chinese Academy of Sciences, Beijing 100190,
China}\affiliation{School of Physical Sciences, University of
Chinese Academy of Sciences, Beijing 100049, China}
\author{Bo Sun}\email[correspondence sent to: ]{sunb@physics.oregonstate.edu }
\affiliation{Department of Physics, Oregon State University,
Corvallis, OR 97331}
\author{Fangfu Ye}
\email[correspondence sent to: ]{fye@iphy.ac.cn}
\affiliation{Beijing National Laboratory for Condensed Matte
Physics and CAS Key Laboratory of Soft Matter Physics, Institute
of Physics, Chinese Academy of Sciences, Beijing 100190, China}
\affiliation{School of Physical Sciences, University of Chinese
Academy of Sciences, Beijing 100049, China}
\author{Yang Jiao} \email[correspondence sent to: ]{yang.jiao.2@asu.edu}
\affiliation{Materials Science and Engineering, Arizona State
University, Tempe, AZ 85287} \affiliation{Department of Physics,
Arizona State University, Tempe, AZ 85287}

\begin{abstract}

Collective cell migration is crucial to many physiological and
pathological processes such as embryo development, wound healing,
and cancer invasion. Recent experimental studies have indicated
that the active traction forces generated by migrating cells in
fibrous extracellular matrix (ECM) can mechanically remodel the
ECM, giving rise to bundle-like meso-structures bridging
individual cells. Such fiber bundles also enable long-range
propagation of cellular forces, leading to correlated migration
dynamics regulated by the mechanical communication among the
cells. Motivated by these experimental discoveries, we develop an
active-particle model with polarized effective attractions (APPA)
for modeling emergent multi-cellular migration dynamics regulated
by ECM-mediated mechanical communications. In particular, the APPA
model generalizes the classic active-Brownian-particle (ABP) model
by imposing a pairwise polarized attractive force between the
particles, which depends on the instantaneous dynamic states of
the particles and mimics the effective mutual pulling between the
cells via the fiber bundle bridge. Active particles with polarized
pairwise attractions exhibit enhanced aggregation behaviors
compared to classic active Brownian particles, especially at lower
particle densities and larger rotational diffusivities.
Importantly, in contrast to the classic ABP system where the
particle velocities are not correlated for all particle densities,
the high-density phase of APPA system exhibits strong dynamic
correlation, which is characterized by the slowly decaying
velocity correlation functions with a correlation length
comparable to the linear size of high-density phase domain (i.e.,
cluster of the particles). The strongly correlated multi-cellular
dynamics predicted by the APPA model are subsequently verified in
{\it in vitro} experiments using MCF-10A cells. Our studies also
indicate the importance of incorporating ECM-mediated mechanical
coupling among the migrating cells for appropriately modeling
emergent multi-cellular dynamics in complex micro-environments.

\end{abstract}



\maketitle

\section{Introduction}







Cell migration in complex extracellular matrix (ECM), a complex
dynamic process involving a series of intra-cellular and
extra-cellular activities \cite{ref12, ref13}, is crucial to many
physiological and pathological processes including tissue
regeneration, immune response and cancer progression \cite{ref1,
ref2, ref3, ref4}. It has been well established that cell
migration can be significantly influenced by the micro-environment
via chemotaxis \cite{ref14}, durotaxis \cite{ref15, ref16,
Brown2009}, haptotaxis \cite{ref17}, and contact guidance
\cite{ref18, ref19, Tranquillo1993}. Migrating cells can generate
active pulling forces via actomyosin contraction \cite{ref_new01,
ref_new02, ref20}, which are transmitted to the ECM fibers through
focal adhesion complexes \cite{ref21, ref22, ref23}. Such active
forces mechanically remodel the local ECM, e.g., by re-orienting
the collagen fibers, forming fiber bundles and increasing the
local stiffness of ECM \cite{ref24, ref25, ref26, ref27, ref28,
ref29, shaohua2019}. Recent studies have indicated that a delicate
balance among the magnitude of the pulling forces, the cell-ECM
adhesion strength, and the ECM rigidity is required to achieve an
optimal mode of single cell migration \cite{ref30}.



\begin{figure*}[ht]
\includegraphics[width=0.85\textwidth,keepaspectratio]{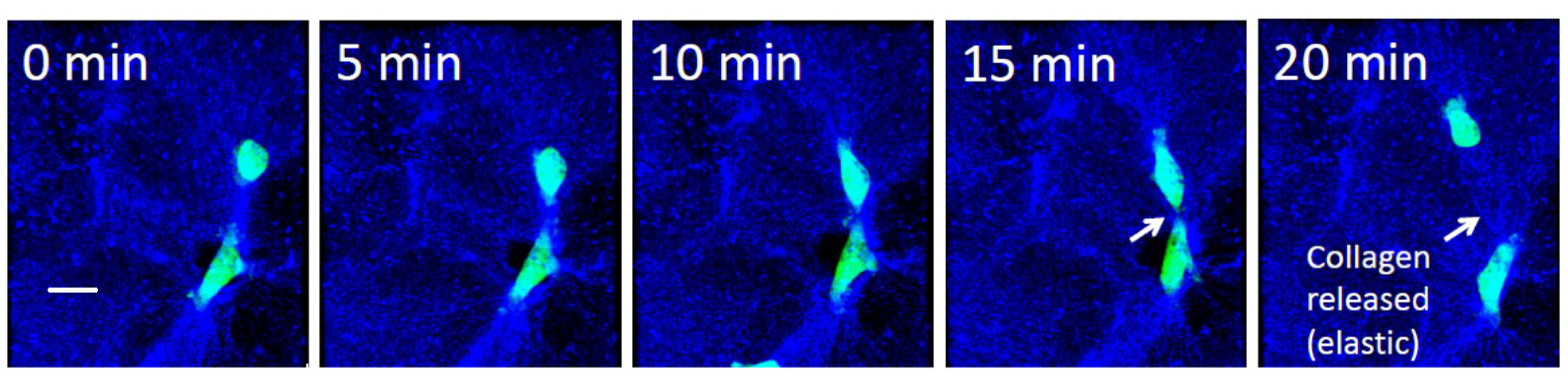}
\label{fig_1}\caption{Confocal microscopy images showing strongly
correlated migration of a pair of MCF-10A cells on thick collagen
gel ($\sim 2 mm$) \cite{qihui2019}. The scale bar is 30 $\mu m$.
The contraction of the cells at the front edge mechanically
remodel the ECM, leading to the formation of fiber bundles
bridging the migrating cells. It has been shown via laser ablation
techniques that the fiber bundles carry significant tensile
forces, and thus strongly suggesting the observed collective
migration dynamics was resulted from the mechanical communications
among the cells via the remodeled fiber bundles. }
\end{figure*}

In a multi-cellular system, the active pulling forces generated by
individual cells can propagate in the ECM and can be sensed by
distant cells \cite{ref5, ref6, ref7, ref8, ref9, ref10, ref11,
ref_new03, ref_new04, ref_new05, ref_new06}. This ECM-mediated
mechanical coupling among the cells could further influence the
migration of the individual cells, which in turn alters the ECM
structure and properties, leading to a rich spectrum of collective
migratory behaviors \cite{ref_new03}. For example, a recent {\it
in vitro} experimental study \cite{qihui2019} showed that highly
motile MCF-10A cells migrating on thick layer of collagen-I gel
develop strongly correlated dynamics via mechanically remodeled
fiber bundles bridging the migrating cells (see Fig. 1). In
addition, it was shown via laser ablation techniques that the
fiber bundles carry significant tensile forces, and thus strongly
suggesting the observed collective migration dynamics was resulted
from the mechanical communications among the cells via the
remodeled fiber bundles \cite{ref_new05}.

Many computational models have been developed to investigate the
individual and multi-cellular migration dynamics \cite{model01,
model02, model03} as well as various sub-cellular processes
involved in cell migration \cite{model04, model05, model06,
model07, model08, model09}. Examples include vertex-based models
\cite{model12}, multi-state cellular Potts models \cite{model13},
cellular automaton models \cite{model14, model15, model16,
model17}. Recently, the influences of the cell-ECM interactions
and/or ECM-mediated indirect cell-cell interactions on individual
and collective migration dynamics are started to be explicitly
considered and incorporated in cell migration models
\cite{new_model01, new_model02, new_model05, model18, ref31,
mike01, mike02, mike03, xiaoming19, new_model06, new_model07,
new_model08}.


The active-particle model is one of most widely used model for
multi-cellular dynamics. In the seminal work of Vicsek and
colleagues \cite{model10}, it was shown that local velocity
correlation among neighboring active particles can induce dynamic
phase transition in the system. The Vicsek model was subsequently
generalized to include cell-cell contact interactions to
investigate collective cellular dynamics such as cell sorting
\cite{model_cellsorting}. More recently, the active Browian
particle model (and its different variants) have received
intensive attentions. In these models, a migrating cell is treated
as an ``active particle'' whose dynamics is mainly determined by
an active self-propelling force, a random drift and various
effective particle-particle and/or particle-environment
interactions \cite{model_lisa, model_ran, model_dijkstra}. A wide
spectrum of collective dynamics have been observed and
investigated in active-particle systems \cite{model11}.

In this work, motivated by the experimental observation that
active pulling forces generated by migrating cells and propagated
via fiber bundles can lead to strongly correlated migration
dynamics (e.g., nearby cells effective attract each other and move
towards each other, see Fig. 1), we develop an active-particle
model with polarized effective attractions (APPA) for modeling
emergent multi-cellular migration dynamics regulated by
ECM-mediated mechanical communications. The APPA model generalizes
the classic active-Brownian-particle (ABP) model by imposing a
pairwise polarized attractive force between the migrating cells
(modeled as active particles), which depends on the instantaneous
dynamic states of cells including the position and velocity
alignment and mimics the effective mutual pulling between the
cells via the fiber bundle bridge.

The APPA model predicts enhanced aggregation behaviors compared to
those predicted by the classic ABP model, especially at lower
particle densities $\phi$ (i.e., the fraction of space covered by
the particles) and larger rotational diffusivities $D_r$ of the
persistent cellular velocity. Importantly, in contrast to the
classic ABP system where the particle velocities are not
correlated for all $\phi$, the high-density phase of APPA system
(i.e., densely packed aggregates of the particles) exhibits strong
dynamic correlation, which is revealed by the velocity vector map
and characterized by the slowly decaying velocity correlation
functions with a correlation length comparable to the linear size
of high-density particle aggregate. The strongly correlated
multi-cellular dynamics predicted by the APPA model are
subsequently verified in {\it in vitro} experiments observing
MCF-10A cells migrating on 3D collagen gels, which exhibit
strongly correlated migration dynamics at high cellular densities.
These results indicate the importance of incorporating
ECM-mediated mechanical coupling among the migrating cells for
appropriately modeling emergent multi-cellular dynamics in complex
micro-environments.

\section{Methods}


\subsection{Active Brownian particle model for single cell migration}


Before describing the details of the active particle model with
polarized effective attractions (APPA) developed here, we first
briefly introduce the classic active Brownian particle (ABP) model
for single cell migration simulations and its biophysical
background. More details and applications of the ABP model and its
generalizations can be found in recent reviews, e.g.,
Ref.\cite{model11}. We note that cell migration in fibrous
extracellular matrix (ECM) is a complex dynamic process involving
a series of intra-cellular and extra-cellular activities including
the development of membrane protrusions, formation of focal
adhesion sites, locomotion due to actin filament contraction, and
detachment of the rear end \cite{ref12, ref13}. In general, the
migration dynamics can be significantly influenced by the
heterogeneity of local ECM microstructure \cite{ref33, ref3333}
and mechanical properties \cite{ref25, MacKintosh05, Safran12,
nat_method15, Frey07, notbom18, notbom19, ref11, ref30}.

In a statistically homogeneous ECM (e.g., one with uniform
collagen density and random fiber orientations), a migrating cell
can be very well modeled as an active Brownian particle, whose
dynamics follows the over-damped Langevin equation, i.e.,
\begin{equation}
\label{eq1} \dot {\bf r}_i(t) = v_0{\bf e}_i + \sqrt{2D_t}{\bf
\Gamma}_i(t)
\end{equation}
where the subscript $i$ is the particle index, $v_0$ is the
persistent speed of the particle, ${\bf e}_i$ is a unitary vector
characterizing the persistent speed direction and subject to
rotational diffusion with diffusion coefficient $D_r$; $D_t$ is
the translational diffusivity of the particle and ${\bf
\Gamma}_i(t)$ is a white noise. The model parameters are
determined by cell phenotype, ECM microenvironment, as well as
stochastic subcellular processes such as actin polymerization,
cell-ECM adhesion turn over. The small set of parameters is
particularly attractive in understanding the complex, multi-scale
processes for 3D cell migration.


The model parameters are determined by cell phenotype, ECM
microenvironment, as well as stochastic subcellular processes such
as actin polymerization, cell-ECM adhesion turn over. The small
set of parameters is particularly attractive in understanding the
complex, multiscale processes for 3D cell migration.





\subsection{Modelling effects of ECM-mediated mechanical coupling via polarized attraction}


A migrating cell also generates active pulling forces
\cite{ref_new01, ref_new02, ref20}, which are transmitted to the
ECM fibers via focal adhesion complexes \cite{ref21, ref22,
ref23}. Such active forces remodel the local ECM, e.g., by
re-orienting the collagen fibers, forming fiber bundles and
increasing the local stiffness of ECM \cite{ref24, ref25, ref26,
ref27, ref28, ref29, shaohua2019}. Importantly, the remodeled
fiber bundles can efficiently transmit the active (pulling) forces
generated by the cells, enabling mechanical dialogs between the
migrating cells.

As illustrated in Fig. 1, the fiber bundles typically formed
between two migrating cells, connecting the polarized pulling ends
of the cells. This allows the pulling forces generate by one cell
to propagate to and sensed by the other cell, and vice versa.
Therefore, we model the effects of this mechanical dialog here
using effective attraction between the cells. Since cell velocity
is typically aligned with the polarization direction, the
effective attraction is also polarized (instead of being
isotropic).We also note that the fiber bundles disappear as the
cell polarization direction changes, e.g., when the two cells are
moving away from one another (see Fig. 1). This suggests that the
fiber bundles, the structural support for the effective
attractions between the cells, are mainly due to elastic ECM
remodeling and temporary \cite{ref_new06, qihui2019}. Therefore,
in our model, we consider the polarized effective attraction only
exist between a pair of migrating cells that are moving {\it
towards} each other (see the mathematical details below). The
underlying ECM network and the remodeled fiber bundles will not be
explicitly considered in our model.

\begin{figure}[ht]
\includegraphics[width=0.35\textwidth,keepaspectratio]{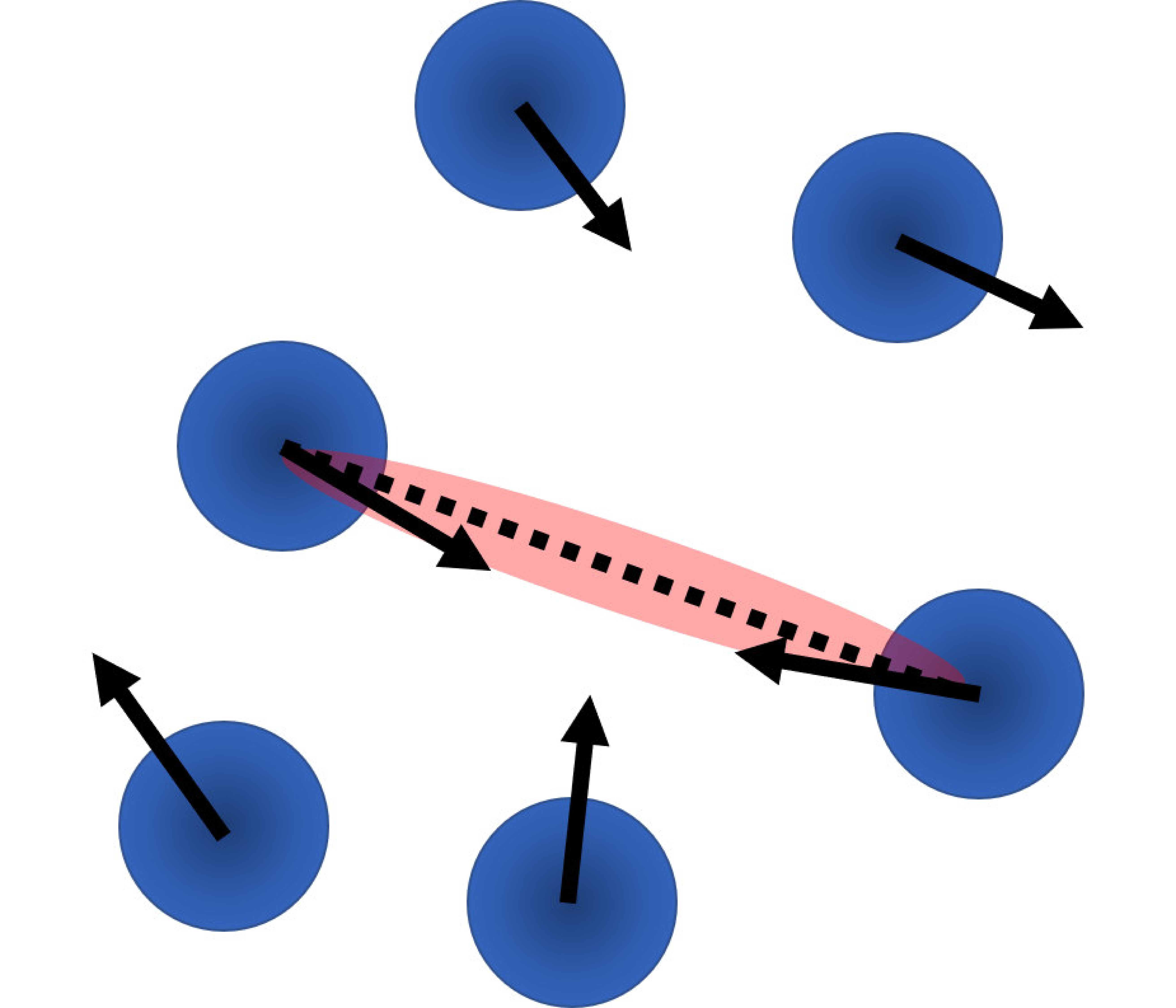}
\label{fig_2}\caption{Schematic illustration of the effective
polarized attraction between two migrating cells, which depends on
both the positions and velocities of the cells. Specifically, the
effective attraction is non-zero only when the two cells are
moving towards one another (i.e., with the velocities
anti-parallel aligned within a prescribed tolerance).}
\end{figure}

Specifically, we consider the dynamics of active particles with
polarized effective attractions are described by the generalized
over-damped Langevin equation, i.e.,
\begin{equation}
\label{eq2} \dot {\bf r}_i(t) = v_0{\bf e}_i + \mu {\bf F}^{ECM}_i
+ \sqrt{2D_t}{\bf \Gamma}_i(t)
\end{equation}
where $\mu$ is the cellular motility and ${\bf F}_i^{ECM}$ is the
total force due to ECM-mediated mechanical coupling among
migrating cells. The other parameters are the same as in Eq.
(\ref{eq1}). We also consider that each particle (cell) possesses
a hard-core exclusion volume modeled as a hard sphere with radius
$R_c$, which prevents the overlapping of the particles upon
contact. The total force ${\bf F}_i^{ECM}$ due to ECM-mediated
mechanical coupling is given by
\begin{equation}
\label{eq_FECM} {\bf F}^{ECM}_i = \sum_j f^a_{ij}(r_{ij}, {\dot
{\bf r}}_i, {\dot {\bf r}}_j)
\end{equation}
where the sum is over all neighboring particles of particle $i$,
and $f^a_{ij}$ is the pair-wise polarized attraction between
particles $i$ and $j$, which depends on the particle separation
distance $r_{ij}$ as well as the dynamic state of the particles
(i.e., ${\dot {\bf r}}_i$, ${\dot {\bf r}}_j$), i.e.,
\begin{equation}
\label{eq_fa} f^a_{ij}(r_{ij}, {\dot {\bf r}}_i, {\dot {\bf r}}_j)
= \left\{{\begin{array}{cc} \epsilon/r^\alpha, & r_{ij}>(R_i + R_j), \\
 & \frac{{\dot {\bf r}}_i \cdot ({\bf r}_j - {\bf r}_i)}{|{\dot {\bf
r}}_i||({\bf r}_j - {\bf r}_i)|}>1-\delta, \\ & \frac{{\dot {\bf
r}}_j \cdot ({\bf r}_i - {\bf r}_j)}{|{\dot {\bf r}}_j||({\bf r}_i
- {\bf r}_j)|}>1-\delta
\\\\ 0, & otherwise
\end{array}}\right.
\end{equation}
where the force parameter $\epsilon$ characterizes the strength of
the effective polarized attraction, which mainly depends on the
cell phenotype and ECM mechanical properties; $R_i$ and $R_j$ are
respectively the radius of particles $i$ and $j$. $\delta$ is a
threshold that quantifies the degree of anti-parallel alignment of
the cell velocities, which is a necessary condition for
ECM-mediated mechanical coupling. The exponent $\alpha$ in the
power-law scaling $1/r^{\alpha}$ characterizes the propagation and
decay of the active force in the ECM. We note that $\alpha = 1$
for an elastic continuum and recent studies indicate that the
decay of active forces in ECM network can be slower (i.e., $\alpha
<1$) due to the unique ``force chain'' structures \cite{ref11}. In
our simulations, the exact value of $\alpha$ does not
significantly influence the collective dynamics and we use $\alpha
= 1$.






Eqs. (\ref{eq2})-(\ref{eq_fa}) are employed in subsequent
simulations of the active particles with polarized attractions
(APPA). In particular, the particles are initially randomly placed
in a periodic square simulation domain without overlapping. The
initial velocities of particles are randomly oriented but the same
magnitude $v_0$. The particles are then evolved according to Eq.
(\ref{eq2}), where the velocities and positions of the particles
are updated at discretized time steps $dt$. The updated velocity
is re-scaled to $v_0$ while keeping its direction. The persistent
direction of each particle is also updated by changing it by a
small angle randomly selected from the interval $[-D_r, D_r]\pi$.
Particle overlaps are removed by pushing each particle back by
half of the overlapping distance along their center-center
direction. The procedure is repeated until a prescribed number of
time steps is achieved and the simulation is terminated.

\subsection{Mean cluster size and velocity correlation function}


We quantify the morphological evolution of the APPA systems using
cluster statistics. Specifically, at a given time step, the
particles are grouped into different clusters. We consider that
two particles belong to the same cluster if the distance between
their centers is less than a threshold (e.g., $d_{ij}<(R_i + R_j +
\delta_d)$). We compute the {\it mean cluster size}
\cite{ref_cluster1, ref_cluster2, ref_cluster3} to quantify the
degree of clustering/aggregation of the particles, i.e.,
\begin{equation}
\label{eq_S} S = \frac{\sum_{k=1}^\infty k^2 n_k
}{\sum_{k=1}^\infty k n_k}
\end{equation}
where $k$ is number of particles that a cluster contains. A
cluster containing $k$ particles is called $k$-cluster. $n_k$ is
the probability that randomly selected cluster in the system is a
$k$-cluster, which is computed by dividing the number of
$k$-cluster by total number of clusters. It can be seen from Eq.
(\ref{eq_S}) that $S \in [0, 1]$. A large $S$ value (e.g., $S \sim
1$) indicates that the majority particles are contained in one (or
a small number of) dominant clusters; while a small $S$ value
indicates that the particles in the system are scatters and the
level of aggregation is low.

In addition, we employ the {\it velocity correlation function} to
quantify the correlated particle dynamics on the two-body (pair)
level, i.e.,
\begin{equation}
\label{eq_Cr} C(r) = <\frac{{\bf u}_i \cdot {\bf u}_j}{|{\bf
u}_i||{\bf u}_j|}>
\end{equation}
where ${\bf u}$ is the instantaneous particle velocities and $|.|$
indicates the magnitude of the vector; $r$ is the cell-center
distance between a pair cells $i$ and $j$; the brackets $<.>$
indicates average over all cell pairs. Perfectly correlated pair
dynamics is associated with $C(r) = 1$. This is the case when the
velocities of all particles are perfectly aligned, e.g.,
indicating the cells are following one another during collective
migration. On the other hand, $C(r) = -1$ is associated with
anti-parallel velocities, i.e., when the cells are moving towards
one another. $C(r) = 0$ indicates no correlations among the
particles (migrating cells).

\section{Results}

\subsection{Enhanced aggregation and dynamic correlation due to
long-range mechanical coupling }







\begin{figure*}[ht]
\centering
\includegraphics[width=0.75\textwidth]{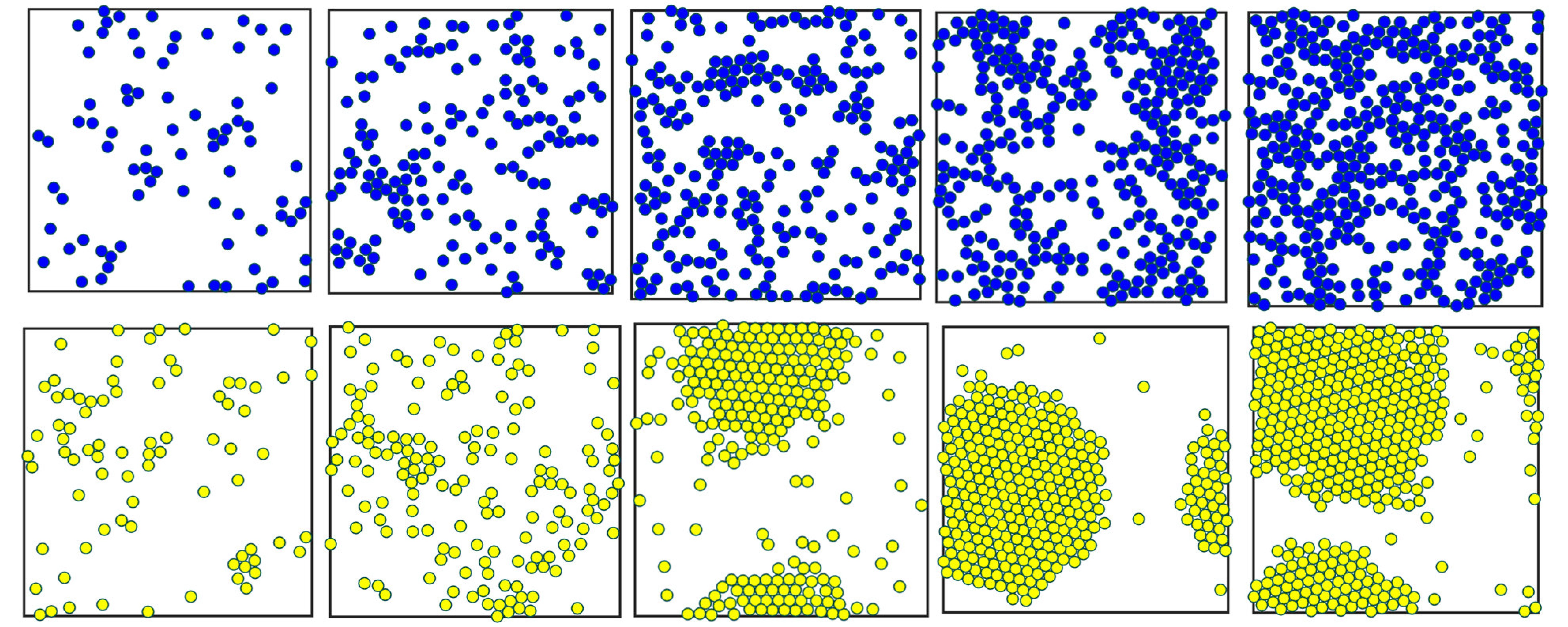}
\caption{Distinct aggregation behaviors at varying particle
densities and large rotational diffusivity $D_r = 0.05$ rising in
the classic ABP systems (upper panels) and in the APPA systems)
(lower panels). The radius of the particles is $R_c = 10 \mu m$
and the linear size of the periodic square simulation domain is $L
= 500 \mu m$. The particle densities from left to right are
respectively $\phi = 0.1, 0.2, 0.3, 0.4$ and 0.5.} \label{fig_3}
\end{figure*}

\begin{figure*}[ht]
\centering
\includegraphics[width=0.75\textwidth]{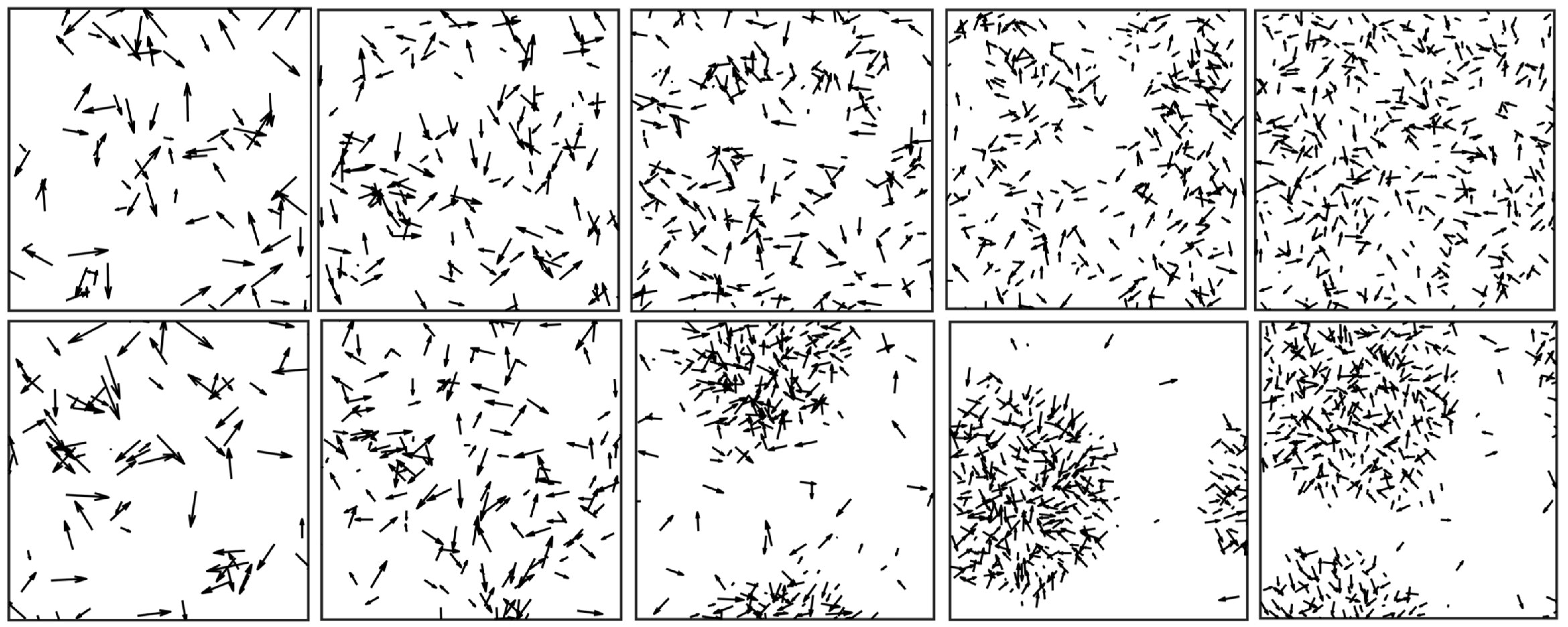}
\caption{Velocity maps (in which the velocity vector associated
with each particle is shown as an arrow at the particle center)
showing distinct velocity correlations at varying particle
densities and large rotational diffusivity $D_r = 0.05$ rising in
the classic ABP systems (upper panels) and in the APPA systems
(lower panels). The particle densities from left to right are
respectively $\phi = 0.1, 0.2, 0.3, 0.4$ and 0.5.} \label{fig_4}
\end{figure*}

\begin{figure*}[ht]
\centering
\includegraphics[width=0.75\textwidth]{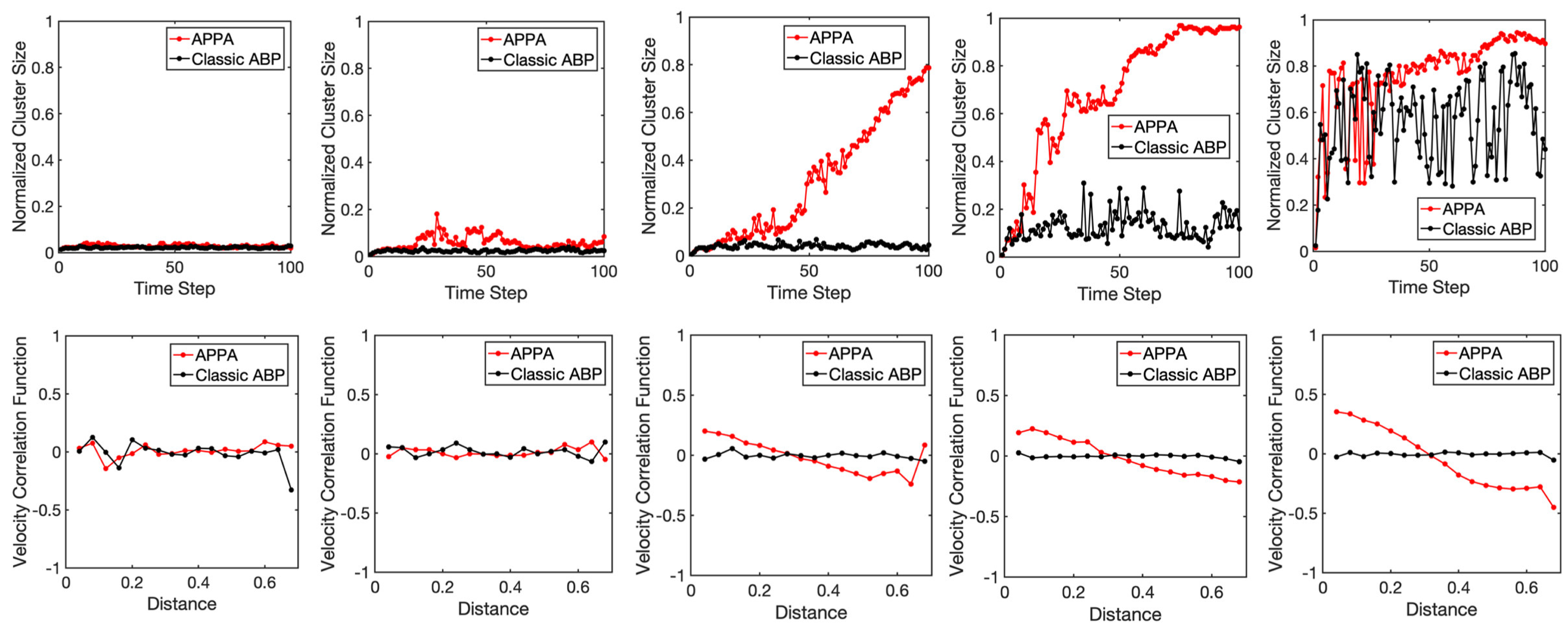}
\caption{Comparison of the cluster statistics $S$ (upper panels,
c.f. Eq.(\ref{eq_S})) and velocity correlation functions $C(r)$
(lower panels, c.f. Eq.(\ref{eq_Cr})) associated with the classic
ABP systems and the APPA systems at large rotational diffusivity
$D_r = 0.05$ and varying particle densities.  The particle
densities from left to right are respectively $\phi = 0.1, 0.2,
0.3, 0.4$ and 0.5. The unit of distance is given by the length of
the square simulation box.} \label{fig_5}
\end{figure*}

We employ the APPA model to investigate the collective dynamics of
multi-cellular systems in homogeneous ECM. Examples of such
systems include the {\it in vitro} experiments where
non-metastatic MCF-10A cells migrate on 3D collagen gel. In this
system, the strong motility of the MCF-10A cells can result in
large active pulling forces, which propagate and influence the
migration of other cells via the remodeled fiber bundles (see Fig.
1). To demonstrate the distinct dynamics resulted from the APPA
model, in particular, the polarized effective attractions that
mimics the mechanical dialogs between the cells, we also simulate
the system using the classic ABP model. We use the following
parameters in our simulations: $v_0 = 0.5 \mu m/min$, $\epsilon =
10 nN/\mu m$, unless otherwise specified. In addition, we set the
translational diffusivity $D_t = 0.01 \mu m^2/min$, which is small
compared to the persistent speed. This is consistent with the
experimental observation that MCF-10A cells on 3D collagen exhibit
strong ballistic-like motions \cite{qihui2019}. We note that the
characteristic dynamics rising in the APPA systems are not
sensitive to the exact values of these parameters.


We first investigate the systems with large rotational diffusivity
(i.e., $D_r = 0.05$). Large $D_r$ values correspond to rapid
relaxing of the persistent velocities, which is known to diminish
particle clustering in classic ABP systems. This can be clearly
seen in Fig. \ref{fig_3} upper panels, which shows the snapshots
of classic ABP system at different particle densities $\phi \in
[0.1, 0.5]$ (i.e., the fraction of simulation box area covered by
the particles). No significant particle clustering occurs until at
very high particle density $\phi = 0.5$. On the other hand, the
APPA system starts to show significant aggregation behavior at
intermediate densities, e.g., $\phi = 0.3$, see Fig. \ref{fig_3}
lower panels. The clustering behaviors of the two different
systems at varying particle densities are also quantified using
time-dependent cluster statistics. Fig. \ref{fig_5} upper panels
show the evolution of mean cluster size $S$ (see Eq. \ref{eq_S}
for definition) in both classic ABP system and APPA system. It can
be seen that $S$ increases rapidly in APPA system with $\phi \ge
0.3$ and asymptotically approaches unity , indicating the
emergence of a dominant cluster containing majority of particles
in the system.


Importantly, the APPA system exists distinct dynamic correlations
compared to the ABP system at high particle densities. This can be
clearly seen from the velocity correlation function $C(r)$ shown
in Fig. \ref{fig_5} lower panels. It can be seen that $C(r)
\approx 0$ for all $r$ and all $\phi$ in the classic ABP system,
regardless of the emergence of particle aggregation, indicating
the particles are not dynamically correlated in the system. In
contrast, the APPA system exhibits strong dynamic correlations,
evidenced by the slow decaying of $C(r)$, which emerges for $\phi
\ge 0.3$, coinciding with the emergence of the particle
aggregation in the corresponding system. To further understand the
emergence of this dynamic correlation, we investigate the particle
velocity distribution in the APPA system. Fig. \ref{fig_4} shows
the velocity maps, in which the velocity vector associated with
each particle is shown as an arrow at the particle center. It can
been seen that the velocities of the particles within the dominant
cluster exhibit strong local and intermediate-ranged alignment
correlation. This distinct dynamics results from the effective
polarized attractions in the APPA system.



\begin{figure*}[ht]
\centering
\includegraphics[width=0.75\textwidth]{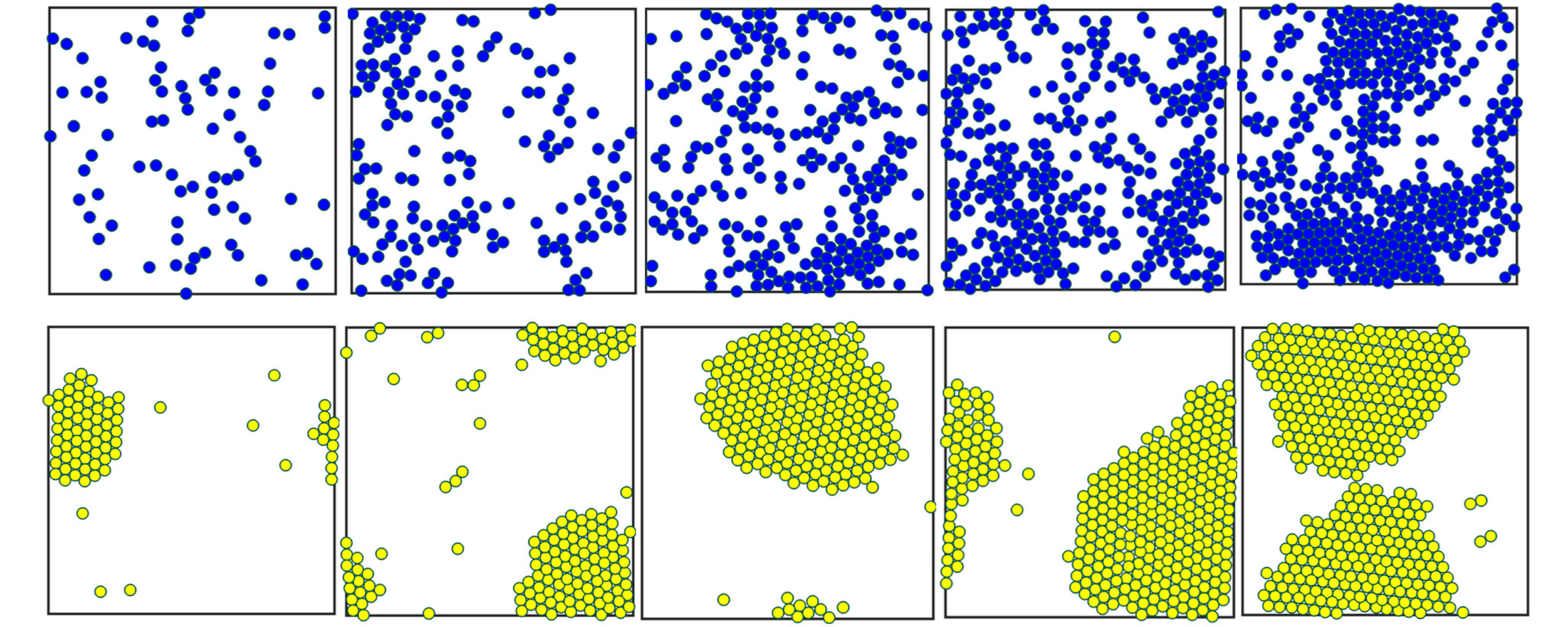}
\caption{Distinct aggregation behaviors at varying particle
densities and small rotational diffusivity $D_r = 0.01$ rising in
the classic ABP systems (upper panels) and in the APPA systems)
(lower panels). The radius of the particles is $R_c = 10 \mu m$
and the linear size of the periodic square simulation domain is $L
= 500 \mu m$. The particle densities from left to right are
respectively $\phi = 0.1, 0.2, 0.3, 0.4$ and 0.5.} \label{fig_6}
\end{figure*}

\begin{figure*}[ht]
\centering
\includegraphics[width=0.75\textwidth]{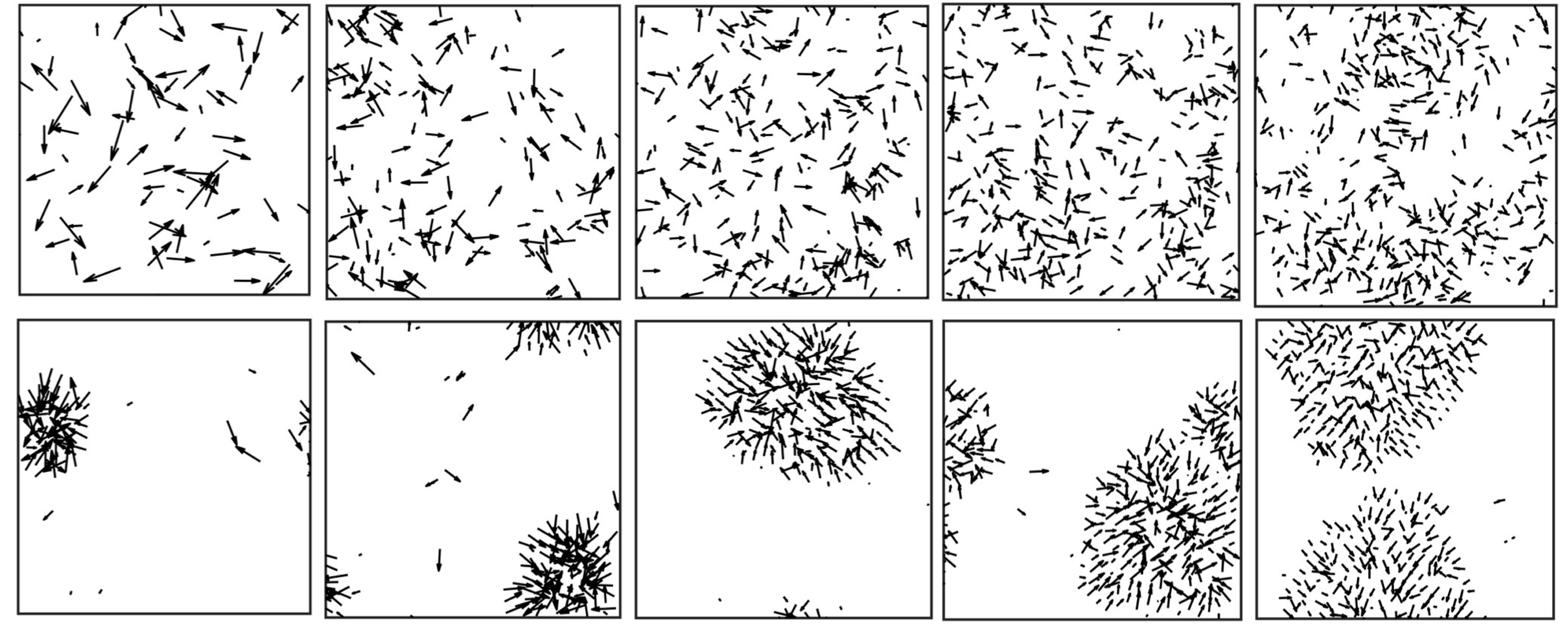}
\caption{Velocity maps (in which the velocity vector associated
with each particle is shown as an arrow at the particle center)
showing distinct velocity correlations at varying particle
densities and small rotational diffusivity $D_r = 0.01$ rising in
the classic ABP systems (upper panels) and in the APPA systems
(lower panels). The particle densities from left to right are
respectively $\phi = 0.1, 0.2, 0.3, 0.4$ and 0.5.} \label{fig_7}
\end{figure*}

\begin{figure*}[ht]
\centering
\includegraphics[width=0.75\textwidth]{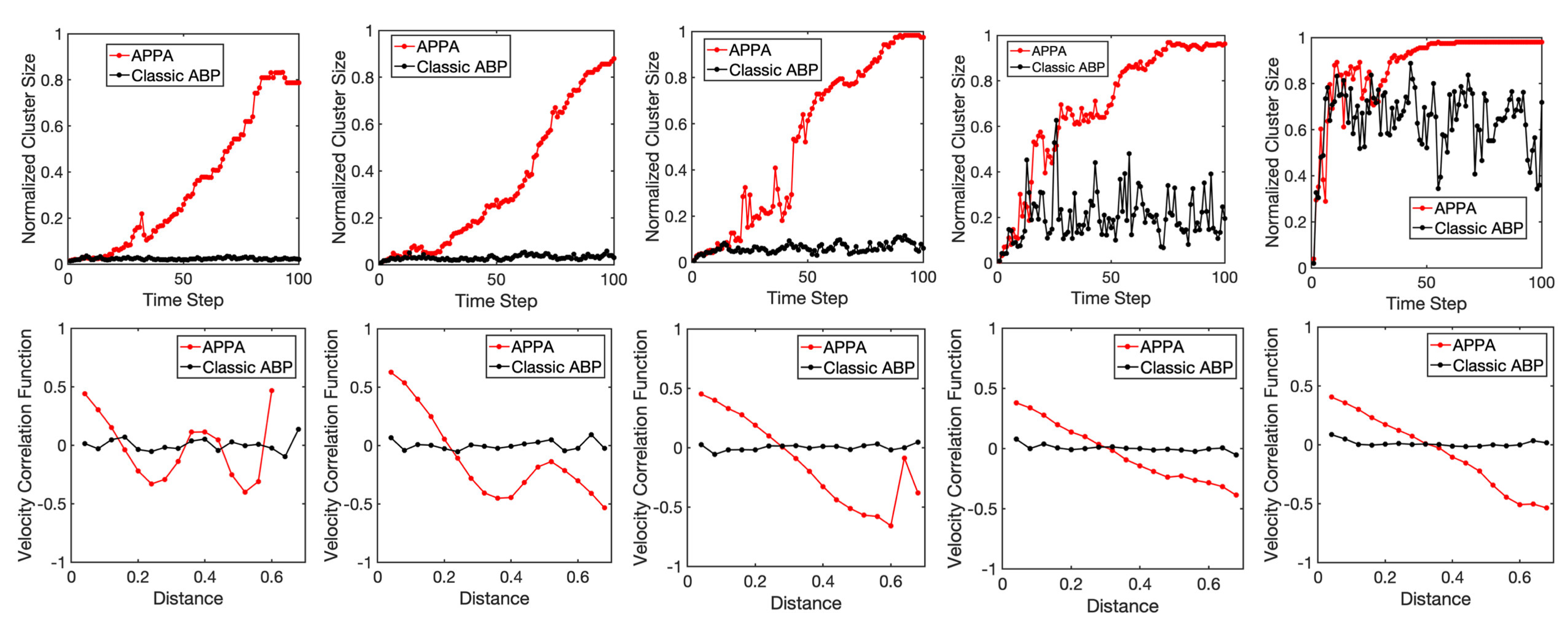}
\caption{Comparison of the cluster statistics $S$ (upper panels,
c.f. Eq.(\ref{eq_S})) and velocity correlation functions $C(r)$
(lower panels, c.f. Eq.(\ref{eq_Cr})) associated with the classic
ABP systems and the APPA systems at small rotational diffusivity
$D_r = 0.01$ and varying particle densities.  The particle
densities from left to right are respectively $\phi = 0.1, 0.2,
0.3, 0.4$ and 0.5. The unit of distance is given by the length of
the square simulation box.} \label{fig_8}
\end{figure*}


We now investigate the systems with small rotational diffusivity
$D_r = 0.01$. It is known that small $D_r$ enhances the clustering
in classic ABP system. Indeed, as shown in Fig. \ref{fig_6} upper
panels, the ABP system starts to show clustering behavior at $\phi
= 0.4$ and significant aggregation is observed at $\phi = 0.5$.
The enhanced clustering behavior is much more significant in the
APPA system, in which single dominant cluster emerges even at the
lowest density $\phi = 0.1$. This is also quantified via the
time-dependent mean cluster size $S$ shown in Fig. \ref{fig_8}
upper panels. Moreover, strong dynamic correlations are observed
in the APPA system for all particle densities $\phi \in [0.1,
0.5]$, which is consistent with the emergence of particle clusters
at low densities. The dynamic correlations can be clearly seen in
the velocity maps (see Fig. \ref{fig_7}) and the corresponding
slow decay in the velocity correlation functions $C(r)$ shown in
Fig. \ref{fig_8} lower panels.

These results indicate that the polarized effective attraction
mimicking the influence of mechanical dialog among the cells can
lead to significantly enhanced dynamics correlations among
particles, especially at lower particle densities and larger
rotational diffusivity. In the next section, we test these model
predictions using {\it in vitro} experiments.

\subsection{Experimental verification via MCF-10A cells on 3D collagen gel}




To test the predictions of APPA model, we observe {\it in vitro}
the migration of multiple MCF-10A cells on 3D collagen I hydrogel
with a collagen concentration $2 mg/ml$. Single cell migration
dynamics is acquired by recording and analyzing cell trajectories
after 12-hours culture. The fibrous microstructure of collagen
gels can support long-range force propagation, which is crucial to
mechanical signaling among the cells. MCF-10A cells are selected
because these cells possess relatively high motility on top of the
collagen gel, but are not able to invade into the gel. On the
other hand, the strong motility of the MCF-10A cells can generate
significant contractile forces during migration, and thus
potentially induce strong cell-ECM mechanical coupling. Additional
details of the experimental studies are reported elsewhere
\cite{qihui2019}.

\begin{figure}[ht]
\centering
\includegraphics[width=0.485\textwidth]{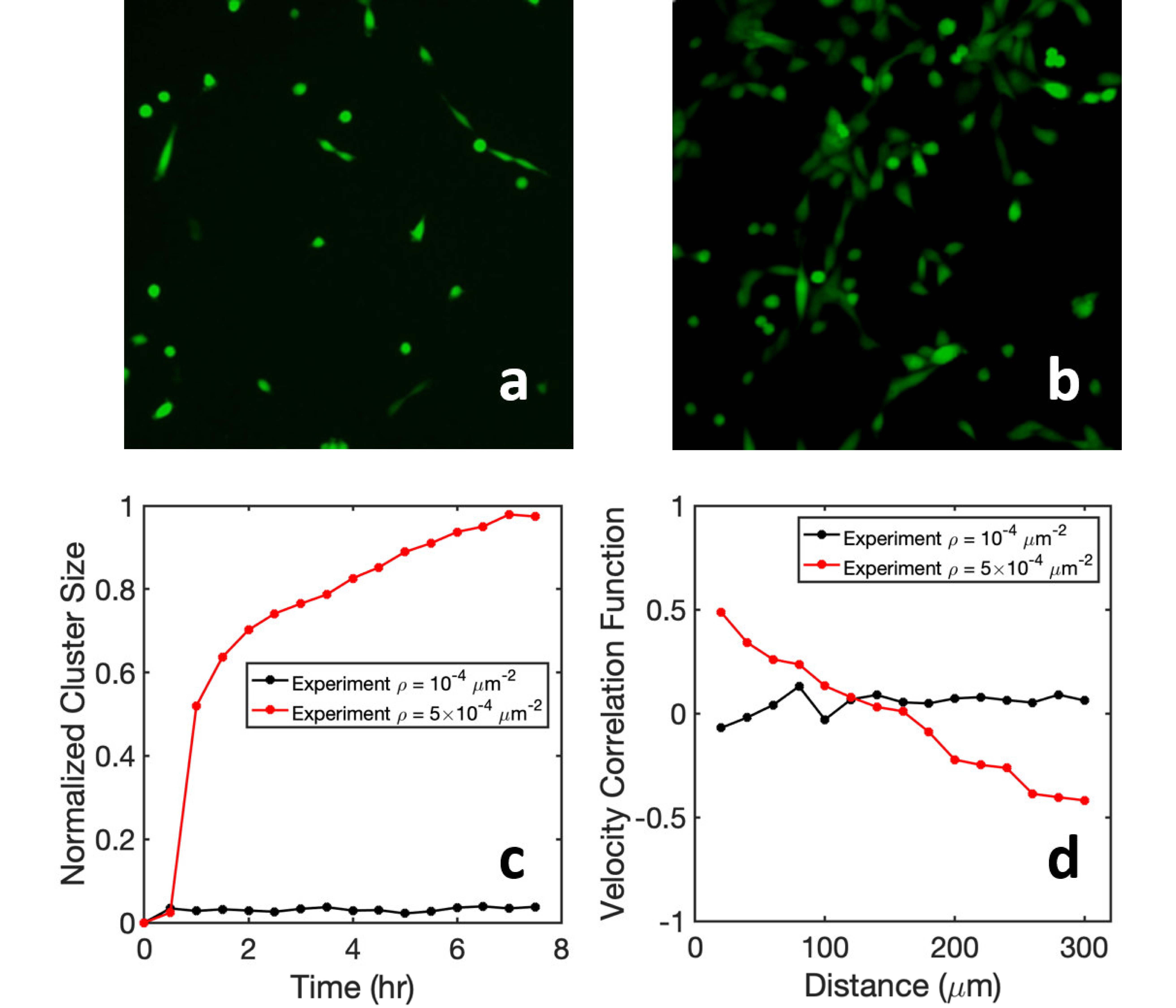}
\caption{Verification of the predicted enhanced aggregation and
dynamic correlation in APPA system using {\it in vitro}
experiments of MCF-10A cells on 3D collagen gels. Panel (a) and
(b) respectively shows snapshots of arrangements of the cells at
$\phi \approx 0.1$ and 0.5 at approximately 6 hours after initial
seeding. The linear size of the image shown here is $\sim 500 \mu
m$. Panel (c) shows the evolution of the mean cluster size $S$
which clearly indicates aggregation behavior at high cellular
density. Panel (d) shows the velocity correlation function $C(r)$,
where the slower decay at high cellular density indicates the
stronger correlated dynamics resulted from the collective
centripetal migration mode.} \label{fig_9}
\end{figure}

In particular, we randomly distribute the MCF-10A cells on
collagen-based ECM with two distinct cell density, corresponding
to the simulation values (i.e., $\phi \approx 0.1$ and 0.5). We
observe rapid and strong aggregation in the high-density system
while the cells in the low-density system remain separated (see
Fig. \ref{fig_9}a and b). We also compute the cluster statistics
$S$ (Fig. \ref{fig_9}c) as well as the velocity correlation
functions $C(r)$ (Fig. \ref{fig_9}d), based on the trajectory data
via cell tracking \cite{qihui2019}. Consistent with the simulation
results, the cells within the cluster in the high-density system
exhibit strong dynamic correlations, as indicated by the
long-range slowly-decaying $C(r)$, as well as the strong
centripetal migration dynamics. The cells in the low-density
system are largely uncorrelated, evidenced by the flat velocity
correlation function and the random cell motions.


\subsection{Phase diagram of active particles with polarized attractions}



The results in previous sections indicate that the polarized
effective attractions may play an important role in giving rise to
unique collective dynamics in multi-cellular systems (such as the
strong dynamic correlation within cell aggregates). In an actual
multi-cell-ECM system, the cell phenotype, ECM microstructure and
physical properties can all affect the active force transmission,
and thus, the effective attraction in our model. In this section,
we systematically vary the key model parameter $\delta$ [c.f. Eq.
(\ref{eq_fa})] that determines the ``region of influence'' of the
polarized forces and investigate its effects on the overall
aggregation behavior and collective dynamics of the system.

\begin{figure*}[ht]
\centering
\includegraphics[width=0.75\textwidth]{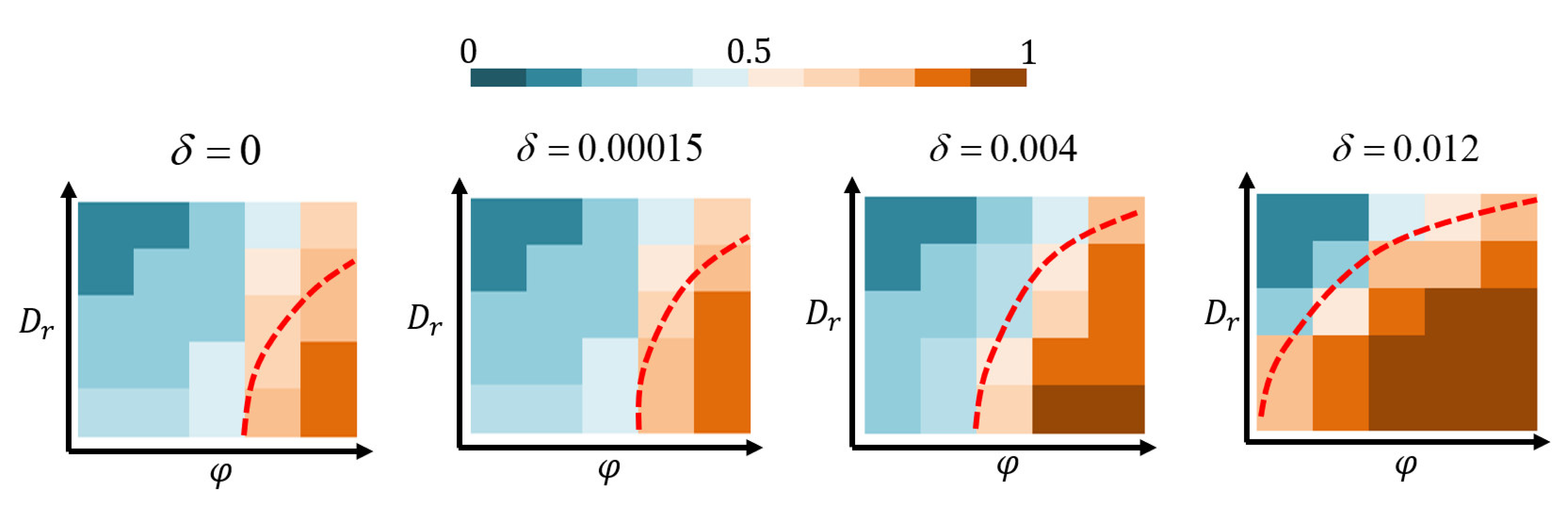}
\caption{Phase diagrams of the APPA system for varying $\delta$,
$D_r$ and $\phi$. Each diagram is associated with a fixed
$\delta$, and collective behaviors of the APPA system at different
rotational diffusivity $D_r \in [0.01, 0.2]$ and particle density
$\phi \in [0.1, 0.5]$ are mapped to two distinct phase regions,
i.e., the aggregation/clustering phase and dispersion/scattered
phase. The colors indicate the values of mean cluster size $S$.
Increasing $\delta$ (i.e., the region of influence of the
polarized attractions) leads to enhanced aggregation/clustering
behavior and thus, stronger dynamic correlation in the aggregation
phase. From the left to the right, $\delta = 0, 0.00015, 0.004$
and 0.012, respectively corresponding to mis-alignment angle
tolerance of $0, \pi/180, \pi/36$ and $\pi/20$. We note that the
classic ABP model is associated with $\delta = 0$.} \label{fig_10}
\end{figure*}


In particular, for each $\delta$, we map the observed collective
behaviors of the APPA system at different rotational diffusivity
$D_r$ and particle density $\phi$ to a ``phase diagram'', where
the aggregation/clustering phase and dispersion/scattered phase
are distinguished. Note that the particles in the aggregation
phase also possess strong dynamic correlations. Figure
\ref{fig_10} shows the phase diagrams of the APPA system for
varying $\delta$, $D_r$ and $\phi$. For a given $\delta$, the
aggregation phase emerges at relatively small $D_r$ and large
$\phi$. We note that the classic ABP model can be obtained by
setting $\delta = 0$. By increasing $\delta$, i.e., force
influence region, the aggregation-dispersion phase boundary is
pushed to larger $D_r$ and smaller $\phi$, indicating enhanced
aggregation behaviors due to mechanical communication. We note
that the largest $\delta$ considered here corresponds to a
mis-alignment angle tolerance of $\pi/20$, which is still
relatively small.





\section{Conclusions and Discussion}





Motivated by recent experimental evidence that ECM-mediate
mechanical coupling among migration cells regulates their
collective dynamics, we develop an active-particle model with
polarized effective attractions (APPA), which generalizes the
classic active-Brownian-particle (ABP) model. Specifically, in the
APPA model, pairwise polarized attractive forces are imposed
between the particles moving towards one another, which mimics the
effective mutual pulling between the cells via the fiber bundle
bridge. The APPA system exhibits enhanced aggregation behaviors
compared to classic ABP system, especially at lower particle
densities $\phi$ and larger rotational diffusivities $D_r$.
Importantly, in contrast to the classic ABP system where the
particle velocities are not correlated for all particle densities,
the high-density phase of APPA system exhibits strong dynamic
correlation, which is characterized by the slowly decaying
velocity correlation functions $C(r)$ with a correlation length
comparable to the linear size of high-density phase domain (i.e.,
cluster of the particles).

We validate our model by accurately reproducing collective
dynamics of MCF-10A breast cancer cells migrating on 3D collagen
gels, including enhanced aggregation behaviors. Moreover, our
model predicts strongly correlated multi-cellular migration
dynamics, which are resulted from the ECM-mediated mechanical
coupling among the migrating cells and also verified in {\it in
vitro} experiments using MCF-10A cells. Our studies indicate the
importance of incorporating ECM-mediated mechanical coupling among
the migrating cells for appropriately modeling emergent
multi-cellular dynamics in complex micro-environments.


Although currently focusing on 2D multi-cellular systems (e.g.,
non-metastatic MCF-10A breast cancer cells migrating on top of 3D
ECM), our model can be easily generalize to investigate the
migration of mesenchymal cells (e.g., invasive MDA-MB-231 breast
cancer cells) in 3D ECM. The key modification is to explicitly
incorporate the effects of ECM degradation by the cells, which
leads to micro-channels that bias cell migration in addition to
the ECM-mediated mechanical coupling. In addition, the effects of
chemotaxis and cell-cell adhesion can also be incorporated into
the model to investigate a wide range of cell lines with different
phenotypes. With proper modifications and generalizations, as well
as efficient parallel implementations, it is expected that the
model could be employed to investigate collective migratory
behaviors and emergent self-organizing multi-cellular patterns
resulted from ECM-mediated mechanical signaling among the cells.



\begin{acknowledgments}
The authors are extremely grateful to the anonymous reviewers. Y.
Z. and Y. J. thank Arizona State University for the generous
start-up funds and the University Graduate Fellowships. Q. F., X.
W., and F. Y. thank the support of the National Natural Science
Foundation of China (Grant Nos.  11704404, 11774394) and the Key
Research Program of Frontier Sciences of Chinese Academy of
Sciences (Grant No. QYZDB-SSW-SYS003). C. Z. E and B. S. thank the
support from the National Science Foundation Grant PHY-1400968.

\end{acknowledgments}


\end{document}